\documentclass{revtex4}
\usepackage{amssymb}
\usepackage{amsmath}

\input{tcilatex}

\begin{document}

\title[ ]{Solutions for f(R) gravity coupled with electromagnetic field }
\author{S. Habib Mazharimousavi}
\email{habib.mazhari@emu.edu.tr}
\author{M. Halilsoy}
\email{mustafa.halilsoy@emu.edu.tr}
\author{T. Tahamtan}
\email{tayabeh.tahamtan@emu.edu.tr}
\affiliation{Department of Physics, Eastern Mediterranean University, G. Magusa, north
Cyprus, Mersin 10, Turkey. }

\begin{abstract}
In the presence of external, linear / nonlinear electromagnetic fields we
integrate $f(R)\sim R+2\alpha \sqrt{R+const.}$ gravity equations. In
contrast to their Einsteinian cousins the obtained black holes are
non-asymptotically flat with a deficit angle. In proper limits we obtain
from our general solution the global monopole solution in $f(R)$ gravity.
The scale symmetry breaking term adopted as the nonlinear electromagnetic
source adjusts the sign of the mass of the resulting black hole to be
physical.
\end{abstract}

\maketitle

\section{\protect\bigskip Introduction}

$f\left( R\right) $ gravity is a modified version of standard Einstein's
gravity which incorporates an arbitrary function of the Ricci scalar ($R$)
instead of the linear one \cite{1} (for a recent review). Depending only on
the Ricci scalar may sound simpler at the initial but the pertinent
nonlinearity makes nothing simpler than the Einstein's gravity with sources.
There are both advantages and disadvantages in adopting such a model. It
contains for instance, its own source known as the curvature source in the
absence of an external matter source. The identification of physical
sources, however, within the nonlinear structure through its equations is
not an easy task at all. For the same reason almost all known solutions,
except very few, result in nonanalytical (i.e. numerical) expressions for
the function $f\left( R\right) $ . Starting from a known function of $%
f\left( R\right) $ a priori is an alternative approach which hosts its own
shortcoming from the outset. Keeping a set of free parameters to be fixed by
observational data can be employed in favour of $f\left( R\right) $ gravity
to explain a number of cosmological phenomena. First of all, to be on the
safe side along with the successes of general relativity most researchers
prefer an ansatz of the form $f\left( R\right) =R+\alpha g\left( R\right) $
, so that with $\alpha \rightarrow 0$ one recovers the Einstein limit. The
struggle now is for the new function $g\left( R\right) $ whose equations are
not easier than those satisfied by $f\left( R\right) $ itself. Without
seeking resort to this latter (and easier) route we have shown recently that 
$f\left( R\right) =\sqrt{R}$ gravity admits exact solution in $6-$%
dimensional spacetime with the external Yang-Mills field \cite{2}. Without
demanding an analytical representation for $f\left( R\right) $ , as a matter
of fact, exact solutions are available in all dimensions with the Yang-Mills
source. Similar results may be investigated with other sources such as the
Maxwell fields. This will be our strategy in the present Letter.

We assume $f\left( R\right) =\xi \left( R+R_{1}\right) +2\alpha \sqrt{R+R_{0}%
}$, in which $\xi ,$ $\alpha ,$ $R_{0}$ and $R_{1}$ are constants, a priori
to secure the Einstein limit by setting the constants $R_{0}=R_{1}=\alpha =0$
and $\xi =1$. This extends a previous study without sources \cite{3} to the
case with sources. Why the square root term in the Lagrangian?. It will be
shown that for $R_{0}=R_{1}=0$ and without external sources such a choice of
square root Lagrangian gives the curvature energy-momentum tensor components
as $T_{t}^{t}=T_{r}^{r},$ $T_{\theta }^{\theta }=T_{\varphi }^{\varphi }=0,$
which signify a global monopole \cite{4}. A global monopole which arises
from spontaneous breaking of gauge symmetry is the minimal structure that
yields non-zero curvature even with zero mass. We test the analogous concept
in $f(R)$ gravity to obtain similar structures. Unlike the case of \cite{2}
our concern here will be restricted to the $4-$dimensional spacetime. As
source, we take electromagnetic fields, both from the linear (Maxwell) and
the nonlinear theories. For the linear Maxwell source we obtain a black hole
solution with electric charge ($Q$) and magnetic charge ($P$) reminiscent of
the Reissner-Nordstrom (RN) solution with different asymptotic behaviors.
That is, our spacetime is non-asymptotically flat with a deficit angle. For
the nonlinear, pure electric source we choose the standard Maxwell invariant
superposed with the square root invariant, i.e. the Lagrangian is given by $%
\mathcal{L}\left( F\right) \sim F+2\beta \sqrt{-F}$, where $F=\frac{1}{4}%
F_{\mu \nu }F^{\mu \nu }$ is the Maxwell invariant and $\beta $ is a
coupling constant. This particular choice has the feature that it breaks the
scale invariance \cite{5} , gives a linear electric potential which plays
role in quark confinement \cite{6,7}. We find out that the scale breaking
parameter $\beta $\ modifies the mass of the black hole. For this reason
Lagrangians supplemented by a square-root Maxwell Lagrangian may find rooms
of applications in black hole physics.

\section{\protect\bigskip $f\left( R\right) $ gravity coupled with Maxwell
field}

The action for $f\left( R\right) $ gravity coupled with Maxwell field in
4-dimensions is given by%
\begin{equation}
S=\int d^{4}x\sqrt{-g}\left[ \frac{f\left( R\right) }{2\kappa }-\frac{1}{%
4\pi }F\right]
\end{equation}%
in which $f\left( R\right) $ is a real function of Ricci scalar $R$ and $F=%
\frac{1}{4}F_{\mu \nu }F^{\mu \nu }$ is the Maxwell invariant. (We choose $%
\kappa =8\pi $ and $G=1$). The Maxwell two-form is chosen to be%
\begin{equation}
\mathbf{F}=\frac{Q}{r^{2}}dt\wedge dr+P\sin \theta d\theta \wedge d\phi
\end{equation}%
in which $Q$ and $P$ are the electric and magnetic charges, respectively.
Our static spherically symmetric metric ansatz is%
\begin{equation}
ds^{2}=-A\left( r\right) dt^{2}+\frac{dr^{2}}{A\left( r\right) }+r^{2}\left(
d\theta ^{2}+\sin ^{2}\theta d\phi ^{2}\right)
\end{equation}%
where $A\left( r\right) $ stands for the only metric function to be found.
The Maxwell equations $\left( \text{i.e. }dF=0=d^{\ast }F\right) $\ are
satisfied and the field equations are given by%
\begin{equation}
f_{R}R_{\mu }^{\nu }+\left( \square f_{R}-\frac{1}{2}f\right) \delta _{\mu
}^{\nu }-\nabla ^{\nu }\nabla _{\mu }f_{R}=\kappa T_{\mu }^{\nu }
\end{equation}%
in which 
\begin{eqnarray}
f_{R} &=&\frac{df\left( R\right) }{dR}, \\
\square f_{R} &=&\frac{1}{\sqrt{-g}}\partial _{\mu }\left( \sqrt{-g}\partial
^{\mu }\right) f_{R}, \\
\nabla ^{\nu }\nabla _{\mu }f_{R} &=&g^{\alpha \nu }\left[ \left(
f_{R}\right) _{,\mu ,\alpha }-\Gamma _{\mu \alpha }^{m}\left( f_{R}\right)
_{,m}\right] ,
\end{eqnarray}%
while the energy momentum tensor is 
\begin{equation}
4\pi T_{\mu }^{\nu }=-F\delta _{\mu }^{\nu }+F_{\mu \lambda }F^{\nu \lambda
}.
\end{equation}%
Furthermore, the trace of the field equation (4) reads%
\begin{equation}
f_{R}R+\left( d-1\right) \square f_{R}-\frac{d}{2}f=\kappa T
\end{equation}%
with $T=T_{\mu }^{\mu }.$ The non-zero energy momentum tensor components are%
\begin{equation}
T_{\mu }^{\nu }=\frac{P^{2}+Q^{2}}{8\pi r^{4}}diag\left[ -1,-1,1,1\right]
\end{equation}%
with zero trace and consequently%
\begin{equation}
f=\frac{1}{2}f_{R}R+3\square f_{R}.
\end{equation}%
One finds%
\begin{eqnarray}
R &=&-\frac{r^{2}A^{\prime \prime }+4rA^{\prime }+2\left( A-1\right) }{r^{2}}%
, \\
R_{t}^{t} &=&R_{r}^{r}=-\frac{1}{2}\frac{rA^{\prime \prime }+2A^{\prime }}{r}%
, \\
R_{\theta }^{\theta } &=&R_{\phi }^{\phi }=-\frac{rA^{\prime }+A-1}{r^{2}}.
\end{eqnarray}%
in which a prime denotes derivative with respect to $r$. Overall, the field
equations read now%
\begin{eqnarray}
f_{R}\left( -\frac{1}{2}\frac{rA^{\prime \prime }+2A^{\prime }}{r}\right)
+\left( \square f_{R}-\frac{1}{2}f\right) -\nabla ^{t}\nabla _{t}f_{R}
&=&\kappa T_{0}^{0}, \\
f_{R}\left( -\frac{1}{2}\frac{rA^{\prime \prime }+2A^{\prime }}{r}\right)
+\left( \square f_{R}-\frac{1}{2}f\right) -\nabla ^{r}\nabla _{r}f_{R}
&=&\kappa T_{1}^{1}, \\
f_{R}\left( -\frac{rA^{\prime }+\left( A-1\right) }{r^{2}}\right) +\left(
\square f_{R}-\frac{1}{2}f\right) -\nabla ^{\theta }\nabla _{\theta }f_{R}
&=&\kappa T_{2}^{2}.
\end{eqnarray}%
Herein 
\begin{equation}
\square f_{R}=A^{\prime }f_{R}^{\prime }+Af_{R}^{\prime \prime }+\frac{2}{r}%
Af_{R}^{\prime },\text{ }\nabla ^{t}\nabla _{t}f_{R}=\frac{1}{2}A^{\prime
}f_{R}^{\prime },\text{ }\nabla ^{r}\nabla _{r}f_{R}=Af_{R}^{\prime \prime }+%
\frac{1}{2}A^{\prime }f_{R}^{\prime },\text{ }\nabla ^{\phi }\nabla _{\phi
}f_{R}=\nabla ^{\theta }\nabla _{\theta }f_{R}=\frac{A}{r}f_{R}^{\prime }
\end{equation}%
and for the details we refer to \cite{2}. The $tt$ and $rr$ components of
the field equations imply%
\begin{equation}
\nabla ^{r}\nabla _{r}f_{R}=\nabla ^{t}\nabla _{t}f_{R}
\end{equation}%
or equivalently%
\begin{equation}
f_{R}^{\prime \prime }=0.
\end{equation}%
This leads to the solution%
\begin{equation}
f_{R}=\xi +\eta r
\end{equation}%
where $\xi $ and $\eta $ are two positive constants \cite{8}. The other
field equations become%
\begin{eqnarray}
f_{R}\left( -\frac{1}{2}\frac{rA^{\prime \prime }+2A^{\prime }}{r}\right) +%
\frac{1}{2}\eta A^{\prime }+\frac{2}{r}A\eta -\frac{1}{2}f &=&\kappa
T_{0}^{0}, \\
f_{R}\left( -\frac{rA^{\prime }+\left( A-1\right) }{r^{2}}\right) +A^{\prime
}\eta +\frac{1}{r}A\eta -\frac{1}{2}f &=&\kappa T_{2}^{2}.
\end{eqnarray}

Now, we make the choice 
\begin{equation}
f\left( R\right) =\xi \left( R+\frac{1}{2}R_{0}\right) +2\alpha \sqrt{R+R_{0}%
}
\end{equation}%
which leads to 
\begin{equation}
R=\frac{\alpha ^{2}}{\eta ^{2}r^{2}}-R_{0}
\end{equation}%
where $\alpha ,$ $R_{0}$ and $\xi $ (from (21)) are constants. As a result
one obtains for $f\left( r\right) $ 
\begin{equation}
f=\frac{\xi \alpha ^{2}}{\eta ^{2}r^{2}}+\frac{2\alpha ^{2}}{\eta r}-\frac{1%
}{2}\xi R_{0}
\end{equation}%
and from (12) we have%
\begin{equation}
-\frac{r^{2}A^{\prime \prime }+4rA^{\prime }+2\left( A-1\right) }{r^{2}}=%
\frac{\alpha ^{2}}{\eta ^{2}r^{2}}-R_{0}.
\end{equation}%
This equation admits a solution for the metric function given by%
\begin{equation}
A\left( r\right) =1-\frac{\alpha ^{2}}{2\eta ^{2}}+\frac{C_{1}}{r}+\frac{%
C_{2}}{r^{2}}+\frac{1}{12}R_{0}r^{2}.
\end{equation}%
Herein the two integration constants $C_{1}$ and $C_{2}$ are identified
through the other field equations (22) and (23) as 
\begin{equation}
C_{1}=\frac{\xi }{3\eta }\text{ and }C_{2}=\frac{\left( Q^{2}+P^{2}\right) }{%
\xi },
\end{equation}%
while for the free parameters we have $\alpha =\eta >0.$ Finally the metric
function becomes%
\begin{equation}
A\left( r\right) =\frac{1}{2}-\frac{m}{r}+\frac{q^{2}}{r^{2}}-\frac{\Lambda
_{eff}}{3}r^{2}
\end{equation}%
where $m=-\frac{\xi }{3\eta }<0,$ $\Lambda _{eff}=\frac{-R_{0}}{4}$ and $%
q^{2}=\frac{\left( Q^{2}+P^{2}\right) }{\xi }.$ The choice of the free
parameters in terms of each other prevents us from obtaining the general
relativity limit, namely the Reissner-Nordstr\"{o}m (RN)-de Sitter (dS)
solution. It is observed that the parameter $\xi $ acts as a scale factor
for mass and charge and for the case $\xi =1$ and $Q=P=0$ the solution
reduces to the known solution given by \cite{3,9}. The properties of this
solution can be summarized as follow: The mass term has the opposite sign to
that of Schwarzschild and the solution is not asymptotically flat, giving
rise to a deficit angle. The latter property is reminiscent of a global
monopole term with a fixed charge. To see the case of a global monopole we
set $R_{0}=0=q^{2}$ (i.e. zero external charges and zero cosmological
constant) and find the energy-momentum components. This reveals that the
non-zero components are $T_{t}^{t}=T_{r}^{r}=-\frac{1}{2r^{2}},$ which
identifies a global monopole \cite{4}. The solution (30) can therefore be
interpreted as an Einstein-Maxwell plus a global monopole solution in $f(R)$
gravity. The area of a sphere of radius $r$ (for $q^{2}=R_{0}=0$) is not $%
4\pi r^{2}$ but $2\pi r^{2}.$ Further, it can be shown easily that the
surface $\theta =\frac{\pi }{2}$ has the geometry of a cone with a deficit
angle $\Delta =\frac{\pi }{2}$ \cite{4}a. It can also be anticipated that a
global monopole modifies perihelion of circular orbits, light bending and
other physical properties. Although in the linear Maxwell theory the sign of
mass is opposite, in the next section we shall show that this can be
overcome by going to the nonlinear electrodynamics with a square root
Lagrangian. Another aspect of the solution is that since $f_{R}>0$ we have
no ghost states.

\section{$f(R)$ Gravity coupled with nonlinear electromagnetism}

\subsection{Solution within nonlinear electrodynamics}

In this section we use an extended model for the Maxwell Lagrangian given in
the action%
\begin{equation}
S=\int d^{4}x\sqrt{-g}\left[ \frac{f\left( R\right) }{2\kappa }+\mathcal{L}%
\left( F\right) \right]
\end{equation}%
where $f\left( R\right) =\xi \left( R+R_{1}\right) +2\alpha \sqrt{R+R_{0}}$,
in which $R_{1}$ and $R_{0}$ are constants to be found while 
\begin{equation}
\mathcal{L}\left( F\right) =-\frac{1}{4\pi }\left( F+2\beta \sqrt{-F}\right)
.
\end{equation}%
Here $\beta $ is a free parameter such that $\lim_{\beta \rightarrow 0}%
\mathcal{L}\left( F\right) =-\frac{1}{4\pi }F$, which is the linear Maxwell
Lagrangian. The main reason for adding this term is to break the scale
invariance and create a mass term \cite{4}. The normal Maxwell action is
known to be invariant under the scale transformation, $x\rightarrow \lambda
x,$ $A_{\mu }\rightarrow \frac{1}{\lambda }A_{\mu },$ ($\lambda =$const.),
while $\sqrt{-F}$ violates this rule. We shall show how a similar term
modifies the mass term in $f(R)$ gravity. Our choice of the Maxwell 2-form
is written as%
\begin{equation}
\mathbf{F}=E\left( r\right) dt\wedge dr
\end{equation}%
and the spherical line element as (3). The nonlinear Maxwell equation reads%
\begin{equation}
d\left( ^{\star }\mathbf{F}\frac{\partial \mathcal{L}}{\partial F}\right) =0
\end{equation}%
which yields the solution%
\begin{equation}
E\left( r\right) =\sqrt{2}\beta +\frac{Q}{r^{2}}
\end{equation}%
with a confining electric potential as $V\left( r\right) =-$ $\sqrt{2}\beta
r+\frac{Q}{r}.$ This is known as the "Cornell potential" for quark
confinement in quantum chromodynamics (QCD) \cite{6,7}. The Einstein
equations implies the same equations as (4-7) and the energy momentum tensor 
\begin{eqnarray}
T_{\mu }^{\nu } &=&\mathcal{L}\left( F\right) \delta _{\mu }^{\nu }-F_{\mu
\lambda }F^{\nu \lambda }\frac{\partial \mathcal{L}}{\partial F}= \\
&&\frac{F}{4\pi }\text{diag}\left[ 1,1,\frac{2\beta }{\sqrt{-F}}-1,\frac{%
2\beta }{\sqrt{-F}}-1\right] ,  \notag
\end{eqnarray}%
with the additional condition that the trace $T_{\mu }^{\mu }=T\neq 0$,
here. Upon substitution into the field equations one gets 
\begin{eqnarray}
R_{1} &=&\frac{4\beta ^{2}}{\xi }+\frac{1}{2}R_{0}. \\
\alpha &=&\eta
\end{eqnarray}%
and a black hole solution results with the metric function 
\begin{equation}
A\left( r\right) =\frac{1}{2}-\frac{4\sqrt{2}\beta Q-\xi }{3\eta r}+\frac{%
Q^{2}}{\xi r^{2}}+\frac{R_{0}}{12}r^{2}.
\end{equation}%
This is equivalent to the solution given in (30) with the same $\Lambda
_{eff}$ but with the new $m=\frac{4\sqrt{2}\beta Q-\xi }{3\eta }$ and $q=%
\frac{Q^{2}}{\xi }.$ This is how the scale breaking term in the Lagrangian
modifies the mass.

For the sake of completeness we comment here that, choosing a magnetic
ansatz for the field two-form as%
\begin{equation}
\mathbf{F}=P\sin \theta d\theta \wedge d\varphi
\end{equation}%
together with a nonlinear Maxwell Lagrangian 
\begin{equation}
\mathcal{L}\left( F\right) =-\frac{1}{4\pi }\left( F+2\beta \sqrt{F}\right)
\end{equation}%
and 
\begin{equation}
R_{1}=\frac{1}{2}R_{0}
\end{equation}%
admits the magnetic version of the solution as 
\begin{equation}
A\left( r\right) =\frac{1}{2}-\frac{4\sqrt{2}\beta P-\xi }{3\eta r}+\frac{%
P^{2}}{\xi r^{2}}+\frac{R_{0}}{12}r^{2}.
\end{equation}%
The magnetic solution, however, is not as interesting as the electric one.

\subsection{Thermodynamical aspects}

The solution we found in the previous section is feasible as far as a
physical solution is concerned. Here we set our parameters, including the
condition $\xi $ and $\eta $ positive, to get $4\sqrt{2}\beta Q-\xi >0$ such
that the solution admits a black hole solution with positive mass as%
\begin{equation}
A\left( r\right) =\frac{1}{2}-\frac{m}{r}+\frac{q^{2}}{r^{2}}+\frac{R_{0}}{12%
}r^{2}.
\end{equation}

Now we wish to discuss some of the thermodynamical properties by using the
Misner-Sharp \cite{2,10} energy to show that the first law of thermodynamics
is satisfied. To do so first we set $R_{0}=0$ and introduce the possible
event horizon as $r=r_{h}$ such that $A\left( r_{h}\right) =0.$ This yields 
\begin{eqnarray}
r_{\pm } &=&m\pm \sqrt{m^{2}-2q^{2}} \\
&&\left( r_{h}=r_{+}\right)  \notag
\end{eqnarray}%
in which 
\begin{equation}
A\left( r\right) =\frac{\left( r-r_{-}\right) \left( r-r_{+}\right) }{2r^{2}}
\end{equation}%
and the constraint $m\geq m_{cri}$ is imposed with $m_{crit}=\sqrt{2}q$ . If
one sets $Q>0$, this condition is satisfied if $Q>\frac{\xi }{\sqrt{2}\left(
4\beta +\frac{3}{\sqrt{\xi }\eta }\right) }$ (providing $4\beta +\frac{3}{%
\sqrt{\xi }\eta }\neq 0$). The choice $m=m_{crit}$ leads to the extremal
black hole. The Hawking temperature is defined as 
\begin{equation}
T_{H}=\frac{A^{\prime }\left( r_{+}\right) }{4\pi }=\frac{r_{+}^{2}-2q^{2}}{%
8\pi r_{+}^{3}}
\end{equation}%
and the entropy \cite{11} 
\begin{equation}
S=\frac{\mathcal{A}_{+}}{4G}\left. f_{R}\right\vert _{r=r_{+}}
\end{equation}%
with $\mathcal{A}_{+}=4\pi r_{+}^{2}$, the surface area of the black hole at
the horizon. The heat capacity of the black hole also is given by 
\begin{equation}
C_{q}=T\left( \frac{dS}{dT}\right) _{q}=-\frac{2}{3}\frac{r_{+}^{2}\pi
\left( 2q^{2}-r_{+}^{2}\right) \left(
12q^{4}+4q^{2}r_{+}^{2}+r_{+}^{4}\right) }{\left( 2q^{2}+r_{+}^{2}\right)
^{2}\left( 6q^{2}-r_{+}^{2}\right) }.
\end{equation}%
which takes both $\left( +\right) $ and $\left( -\right) $ values. Both the
vanishing / diverging $C_{q}$ values indicate special points at which the
system attains thermodynamical phrase changes.

The first law of thermodynamics can be written as 
\begin{equation}
TdS-dE=PdV
\end{equation}%
in which%
\begin{equation}
dE=\frac{1}{2\kappa }\left[ \frac{2}{r_{h}^{2}}f_{R}+\left( f-Rf_{R}\right) %
\right] \mathcal{A}_{+}dr_{+}
\end{equation}%
with $E$ the Misner-Sharp energy and $T=\frac{A^{\prime }}{4\pi }$ the
Hawking temperature. Further, $S=\frac{\mathcal{A}_{+}}{4}f_{R}$ stands for
the black hole entropy, $p=T_{r}^{r}=T_{0}^{0}$ \ is the radial pressure of
matter fields at the horizon and finally the change of volume of the black
hole at the horizon is given by $dV=\mathcal{A}_{+}dr_{+}.$ One can easily
show that the first law in thermodynamics in the form introduced above is
satisfied.

\section{Conclusion}

Exact solutions for nowadays popular, modified gravity model known as $f(R)$
gravity with external sources (i.e. $T_{\mu \nu }^{matter}\neq 0$) are rare
in the literature. We attempt to fill this vacuum partially by considering
external electromagnetic fields (both linear and nonlinear) in $f(R)$
gravity with the ansatz $f\left( R\right) =\xi \left( R+R_{1}\right)
+2\alpha \sqrt{R+R_{0}}$. In this choice $R_{0}$\ is a constant related to
the cosmological constant, the constant $R_{1}$ is related to $R_{0}$ while $%
\alpha $ is the coupling constant for the correction term. This covers both
the cases of linear Maxwell and a special case of power-law nonlinear
electromagnetism. The non-asymptotically flat black hole solution obtained
for the Maxwell source is naturally different and has no limit of the RN
black hole solution. In the limit of $Q=P=\Lambda _{eff}=0$ we obtain the
metric for a global monopole in $f(R)$ gravity. Our solution can
appropriately be interpreted as a global monopole solution in the presence
of the electromagnetic fields. The thermodynamical properties of our black
hole solution is analyzed by making use of the Misner-Sharp formalism and
shown to obey the first law. As the nonlinear electromagnetic Lagrangian we
choose the normal Maxwell, supplemented with the square root Maxwell
invariant which amounts to a linear electric field. This latter form is
known to break the scale invariance yielding a linear potential which is
believed to play role in quark confinement problem. Within $f(R)$ gravity
the presence of scale breaking term modifies the mass of the resulting black
hole. The advantage of employing square-root Maxwell Lagrangian as a
nonlinear correction can be stated as follows: Beside confinement in the
linear Maxwell case we have in $f(R)$ gravity an opposite mass term while
with the coupling of the aquare-root Maxwell Lagrangian we can rectify the
sign of this term.

\end{document}